\documentclass[preprint,12pt]{elsarticle}

\usepackage{amssymb}
\usepackage{hyperref}

\usepackage{setspace}

\usepackage{lineno}

\usepackage{footmisc}

\biboptions{round, authoryear}

\journal{Icarus}

\usepackage{adjustbox}
\usepackage{array}

\newcolumntype{R}[2]{%
    >{\adjustbox{angle=#1,lap=\width-(#2)}\bgroup}%
    l%
    <{\egroup}%
}

\begin{document}

\begin{frontmatter}



\title{Cascade disruptions in asteroid clusters}

\author[af1,af2]{Petr Fatka}
\author[af1]{Petr Pravec}
\author[af2]{David Vokrouhlick\'y}

\address[af1]{Astronomical Institute, Academy of Sciences of the Czech Republic, Fri\v{c}ova 1, Ond\v{r}ejov CZ-25165, Czech Republic}
\address[af2]{Institute of Astronomy, Charles University, Prague, V Hole\v{s}ovi\v{c}k\'{a}ch 2, CZ-18000 Prague 8, Czech Republic}

\begin{abstract}
We studied asteroid clusters suggesting a possibility of at least two disruption events in their recent history ($\leq 5$ Myr). We searched for new members of known asteroid pairs and clusters and we verified their membership using backward orbital integrations. We found four asteroid clusters, namely the clusters of (11842) Kap'bos, (14627) Emilkowalski, (63440) 2001~MD30 and (157123) 2004~NW5 that show at least two secondary separation events that occurred at significantly different times. We considered a possible formation mechanism for these clusters: The parent of an asteroid  cluster was spun up to its critical rotation frequency, underwent a rotation fission and was slowed down by escape of the newly formed secondary/ies. Then the YORP effect spun up the primary again and it reached its critical rotation frequency and underwent another fission. We created a simple model to test whether the scenario of two rotation fission events of a parent primary induced via the YORP effect is possible for the four clusters. We obtained a good agreement between the model and the cluster properties for the clusters of Kap'bos and (63440). For the cluster of Emilkowalski, our model explained the unusually slow rotation of the primary. However, the time needed for the primary to reach its critical frequency after the first fission event was predicted to be too long by a factor of several. We suspect, considering also its D type taxonomic classification and the existence of a dust band associated with the cluster, that the asteroid Emilkowalski may actually be a cometary nucleus. Regarding the cluster of (157123), the final rotational frequency of the primary after the last fission event predicted by our model is in a good agreement with the observed rotation frequency of (157123). However, a separation of the older secondary is not possible in our model due to the deficiency of free energy needed for an escape of the large secondary. This could be due to an error in the $H$ value of the secondary or the possibility that we did not find the real primary of this cluster. 

\end{abstract}


\end{frontmatter}


\section{Introduction}
The existence of young asteroid clusters on highly similar heliocentric orbits is known for over a decade now. The first four of these very young systems (with ages $<$ a~few million years) were found by \citet{Nesvorny:2006a} and \citet{Nesvorny:2006b}. Specifically, the cluster of (1270) Datura was identified in the first publication. In the later paper, the authors identified three new asteroid clusters, namely the clusters of (14627) Emilkowalski, (16598) Brugmansia and (21509) Lucascavin. All of these four asteroid clusters have ages $\le 1$ Myr. Another five clusters were discovered by \citet{Pravec:2009} as a by-product of their asteroid pair analysis, namely the clusters of (6825) Irvine, (10321) Rampo, (18777) Hobson, (39991) Iochroma and (81337) 2000 GP36\footnote{\citet{Vokrouhlicky:2011} found that it is a part of the larger cluster of (2384) Schulhof. See also \citet{Vokrouhlicky:2016a}.}. Their initial age estimation showed that all the five clusters are younger than 2 Myr. Later \citet{Novakovic:2014} found the cluster of (20674) 1999 VT1 and estimated its age to about 1.5 Myr.

In general, the orbits of asteroid cluster members disperse in time due to the perturbations from the major planets and the orbital drift by the Yarkovsky effect and they mix in the space of orbital elements with background asteroids relatively fast. The rate of the orbit dispersion depends on several factors (e.g., the local density of surrounding background asteroid orbits, or the influence of nearby resonances), but even in the stable and sparsely populated regions of the main belt, it is difficult to identify an asteroid cluster after only a few million years since its formation. The difficulty of cluster identification increases with smaller number of its members. Nine of the 10 clusters mentioned above were found by analyzing the osculating orbital elements of asteroids, only the cluster of (20674) was identified in the space of proper orbital elements \citep[e.g.,][]{Knezevic:2002}. In our recent work \citet{Pravec:2018} we analyzed asteroid mean orbital elements \citep[e.g.,]{Milani:1998} for an improved asteroid cluster search. Mean orbital elements are obtained from osculating ones by removing the short-period perturbations, therefore the analysis was more robust and independent of the current phase of the orbital element oscillations at present epoch. In that work, we found several new members of the previously known clusters as well as three new clusters, namely the clusters of (11842) Kap'bos, (22280) Mandragora and (66583) Nicandra.

\citet{Vokrouhlicky:2008b} proposed that at least some of the 60 asteroid pairs they identified were formed by Yarkovsky-O'Keefe-Radzievski-Paddack (YORP) effect induced spin-up \citep[see][and references therein]{Vokrouhlicky:2015} and subsequent rotational fission of rubble pile asteroids. In principle, asteroid clusters formed by rotational fission and asteroid families created by catastrophic collisions can by distinguished by comparing relative velocities of their members at the moment when the system became unbound. In the case of a cluster formed by rotational fission, the relative velocity of an escaping secondary to the primary\footnote{The term ``primary'' is used for the largest body of a cluster. The term ``secondary'' is used for any smaller member of given cluster.} is expected to be close to the escape velocity from the surface of the primary asteroid \citep{Scheeres:2007, Pravec:2010}, which is typically less than a few meters per second for observed asteroid cluster primaries.

In our previous work \citet{Pravec:2018} we found three new members of the asteroid cluster of (14627) Emilkowalski. These new members are close to each other in the space of mean orbital elements, but are more distant from the primary than the previously known members, therefore they were not treated as member candidates before (see more details in Section 4.2). It was very intriguing to find out that these new members showed past convergence to the primary, but at significantly different times than two of the three previously known members, suggesting a cascade disruption. Being motivated by that interesting finding, in this work we search for another cases of asteroid clusters with multiple separation events and we perform detailed analysis of four such interesting clusters that we found.

\section{Cascade disruption candidate selection}
We started our search for asteroid clusters with multiple separation times with checking the already known asteroid clusters and pairs for new members. For description of similarity of two asteroid orbits we used a metric used by \citet{Vokrouhlicky:2008b}, where a distance $d_{\mathrm{osc/mean}}$ of two orbits in the five-dimensional space of (osculating or mean) orbital elements ($a, e, i, \Omega, \varpi$) is defined as

\begin{equation}
\left( \frac{d_{\mathrm{osc/mean}}}{na} \right)^2 = k_a \left( \frac{\delta a}{a} \right)^2 + k_e \left( \delta e \right)^2 + k_i \left( \delta \sin i \right)^2 + k_{\Omega} \left( \delta \Omega \right)^2 + k_{\varpi} \left( \delta \varpi \right)^2
\end{equation}
where $n$ is the mean motion, ($\delta a, \delta e, \delta \sin i, \delta \Omega, \delta \varpi$) is the separation vector of the orbits and the coefficients are $k_a = 5/4, k_e = k_i~=~2, k_{\Omega} = k_{\varpi} = 10^{-4}$. We used the catalog of mean orbital elements downloaded from the \texttt{AstDyS-2} website\footnote{https://newton.spacedys.com/astdys/}.

We started our search with the 13 asteroid clusters from \citet{Pravec:2018} and the 3 newly discovered clusters from \citet{Pravec:2019}. For each cluster, we selected asteroids with distances in the space of mean orbital elements $d_{\mathrm{mean}} \le 400-750$ m/s, depending on the local density of background asteroid orbits. The lower limit was used in densely populated regions and the upper limit was used in sparsely populated regions of the main asteroid belt. We typically found $150 - 350$ asteroids within the chosen distance from the primary of the studied cluster.

For a given cluster and the $150 - 350$ found nearby asteroids, we performed backward integrations of their nominal orbits obtained from the \texttt{AstDyS-2} website for epoch MJD 58400.0. We took into account only gravitational interactions\footnote{In our integrations we included the Sun, the eight major planets, Pluto, Ceres, Vesta and Pallas.} and propagated the asteroid orbits 5 Myr into the past. We recorded the time evolution of secular angles $\Omega \left( t \right)$ and $\varpi \left( t \right)$ and the time evolution of distance $d_{\mathrm{osc}} \left( t \right)$ between the primary asteroid (orbital elements with subscript $\mathrm{prim}$) and each of the tested asteroids (orbital elements with index $i$). We searched for cases where the differences of both secular angles $\Delta \Omega \left( t \right) \equiv \Omega_{i} \left( t \right) - \Omega_{\mathrm{prim}} \left( t \right)$ and  $\Delta \varpi \left( t \right) \equiv \varpi_{i} \left( t \right) - \varpi_{\mathrm{prim}} \left( t \right)$ were close to zero at about the same time and also the distance $d_{\mathrm{osc}} \left( t \right)$ was reasonably small. Because we neglected the Yarkovsky effect and used only the asteroid nominal orbits in this initial analysis, we used rather relaxed limits for the selection of member candidates, specifically, a minimum differences in the secular angles up to several degrees and distances $d_{\mathrm{osc}} \left( t \right)$ up to $\sim 70$ m/s.  This allowed us to find potential members of each studied cluster that were more distant from its primary component than the previously known members and that are already mixing with the background population of asteroids\footnote{We also applied this procedure to the 93 asteroid pairs studied in \citet{Pravec:2019}, with a difference that we tested only 100 asteroids closest to the primary of each pair.}. 

Filtering member candidates from unrelated background asteroids was done by manual inspection of secular angles evolution, which was a rather easy job for a single person to perform the task. Specifically, we evaluated whether $\Delta \Omega$ and $\Delta \varpi $ were close to zero at about same time. We also checked if $d_{\mathrm{osc}}$ was reasonably small around that time. We also assessed the overall smoothness of the $\Delta \Omega$ and $\Delta \varpi $ time evolution, whether it was disturbed by some resonance or a close approach to a planet or  massive asteroid. The idea behind this method is very similar to the one used by \citet{Novakovic:2012} that they named \textit{Selective Backward Integration Method}, which helps to discriminate between real cluster members and background asteroids.

Finally, we also analyzed the full catalog of the mean orbital elements of asteroids, but we did not find any new candidate clusters for a cascade fission.



\section{Membership confirmation and age estimation}
After we chose the candidate clusters with possible two (or more) secondary separation events, we performed orbital clone integrations for each member of a given cluster into the past. The goal of these backward integrations was to verify whether a slow and close encounter between the primary and the tested secondary occurred in the past and to estimate the time of their separation. In two cases we also searched for potential slow and close encounters between two selected secondaries (see Sections 4.1 and 4.2) to check whether the smaller of the two secondaries could possibly separate from the larger secondary. 

For each cluster member, we created 1000 geometric clones, which represented different realizations of its orbit. These geometric clones were created in the six-dimensional space of equinoctial orbital elements \textbf{E} using the probability distribution $p\left(\textbf{E}\right) \propto \exp \left( -\frac{1}{2} \Delta \textbf{E} \cdot \Sigma \cdot \Delta \textbf{E} \right)$, where $\Delta \textbf{E} = \textbf{E} - \textbf{E}^{*}$ is the difference with respect to the best-fit orbital values $\textbf{E}^{*}$ and $\Sigma$ is the normal matrix of the orbital solution  \citep{Milani:2010}. Each of these geometric clones was assigned with a different strength of the Yarkovsky effect, which acted on a given clone in the form of a fake transverse acceleration with a chosen magnitude providing secular change in the semi-major axis $\dot{a}_{\mathrm{Yark}}$ \citep{Farnocchia:2013}. It was randomly chosen from the range $\left\langle -\dot{a}_{\mathrm{max}}, \dot{a}_{\mathrm{max}} \right\rangle$, where $\dot{a}_{\mathrm{max}}$ was estimated from the asteroid size\footnote{We used the relation $D = \frac{1329}{\sqrt{p_v}} 10^{-H/5}$ from \citet{Harris:1997} to estimate asteroid diameters.} \citep{Vokrouhlicky:1999}. These limit values for the semi-major axis drift rate correspond to bodies with zero obliquity, for which the diurnal variant of the effect is optimized, and the diurnal thermal parameter equal to the square root of two, for which the magnitude of the Yarkovsky effect is maximal. Earlier versions of this method were used in, e.g., \citet{Pravec:2010, Pravec:2018}.

The requirement for a close encounter between the orbital clones of tested asteroids is obvious - a parent asteroid was split into two asteroids so they were physically close one to each other at that time. And since the secondary cannot temporarily orbit around the primary at distances greater than the radius of the Hill sphere\footnote{$R_{\mathrm{Hill}} \sim a D_1 \frac{1}{2} \left( \frac{4\pi}{9}\frac{G\rho_1}{\mu} \right)^{1/3}$, where $a$ is the semi-major axis of the primary's heliocentric orbit, $D_1$ is its diameter, $\rho_1$ is its bulk density, $G$ is the gravitational constant and $\mu$ is the gravitational parameter of the Sun (Pravec et al., 2010, Supplementary Information).} $R_{\mathrm{Hill}}$ of the primary, the secondary must escape at distance comparable to $R_{\mathrm{Hill}}$, which is typically a few hundred kilometers for the studied asteroids. Since the theory of rotational fission of a rubble pile asteroid (Scheeres, 2007; Pravec et al., 2010) predicts a gentle escape of a secondary at relative velocities $v_{\mathrm{rel}}$ comparable with the escape velocity\footnote{$v_{\mathrm{esc}}  \sim D_1 \frac{1}{2} \left( \frac{8 \pi}{3} G \rho_1 \right)^{1/2}$ (Pravec et al., 2010, Supplementary Information).} $v_{\mathrm{esc}}$ from the primary's surface, we require the relative velocity of the two clones during an encounter to be similarly low. With this requirement, we filtered out random orbital clone encounters at high relative velocities that are not relevant for the actual separation event of the secondary. Because of the uncertainty of initial orbital elements, the finite number of used orbital clones and the limited precision of an integrator, we relaxed the limits of what we consider to be a close and slow encounter somewhat beside the limits suggested by the rotational fission theory. We chose following limits for the relative distance $r_{\mathrm{rel}}$ and relative velocity between the clones $r_{\mathrm{rel}} \le 10 $ or $15 R_{\mathrm{Hill}}$ and $v_{\mathrm{rel}} \le 1,2 $ or $ 4 v_{\mathrm{esc}}$. The more strict limits were used in cases of younger ages or if the studied asteroids were located in non-chaotic zones of the main asteroid belt, while the loosened limits were typically used in cases with estimated separation times $> 1$ Myr and in cases with the orbits affected by some orbital chaoticity.

We used the fast and accurate implementation of a Wisdom-Holman symplectic integrator \texttt{WHFast} \citep{Rein:2015} from the \texttt{REBOUND} package \citep{Rein:2012} and added the Yarkovsky effect described above. We accounted for the gravitational attraction of the Sun, the 8 major planets, two dwarf planets Pluto and Ceres and two large asteroids Vesta and Pallas. Every tenth day of the integration time we checked all the clone combinations ($1000 \times 1000$) of the primary and the secondary, calculated their $r_{\mathrm{rel}}$ and $v_{\mathrm{rel}}$ and saved the encounter if it satisfied the set limits. If the encounter of two clones was recorded several times in a row (i.e., the encounter lasted $> 10$ days), we picked only the record with the smallest $v_{\mathrm{rel}}$.

To estimate the separation time (age) $T_{\mathrm{sep}}$ from the recorded encounters, we used the median (i.e., the 50th percentile) value of the distribution as a nominal estimate. For an uncertainty estimation of the separation time, we adopted the 5th and the 95th percentile of the distribution for the lower and the upper limit on the separation time, respectively.

We also calculated a probability that the most distant\footnote{Based on the distances in the space of mean orbital elements.} member of each cluster was falsely identified as a cluster member (this probability is always lower for other cluster members that are closer to the primary). We followed and adapted the approach by \citet{Pravec:2009}. This probability $p_1$ is given by Poisson statistics $p_1\left(V\right) = \nu e ^{-\nu}$, where $\nu=\eta V$ and $\eta$ is the number density within volume $V$. For $\nu \ll 1$, the formula becomes simply $p_1\left(V\right) = \nu$. To calculate $\nu$ for specific cluster members, we employ $R_0$, which is the radius of a hypersphere with specific volume\footnote{$R_0$ is a characteristic distance of objects for the observed density $\eta$.} and we obtain $\nu = \eta V = \left(d_{\rm{mean}}/R_0 \right)^5$. As we show below, the probabilities of that the identified cluster members are interlopers is extremely low.

\section{Individual clusters}
\begin{figure}[h!]
\centering\includegraphics[width=\linewidth]{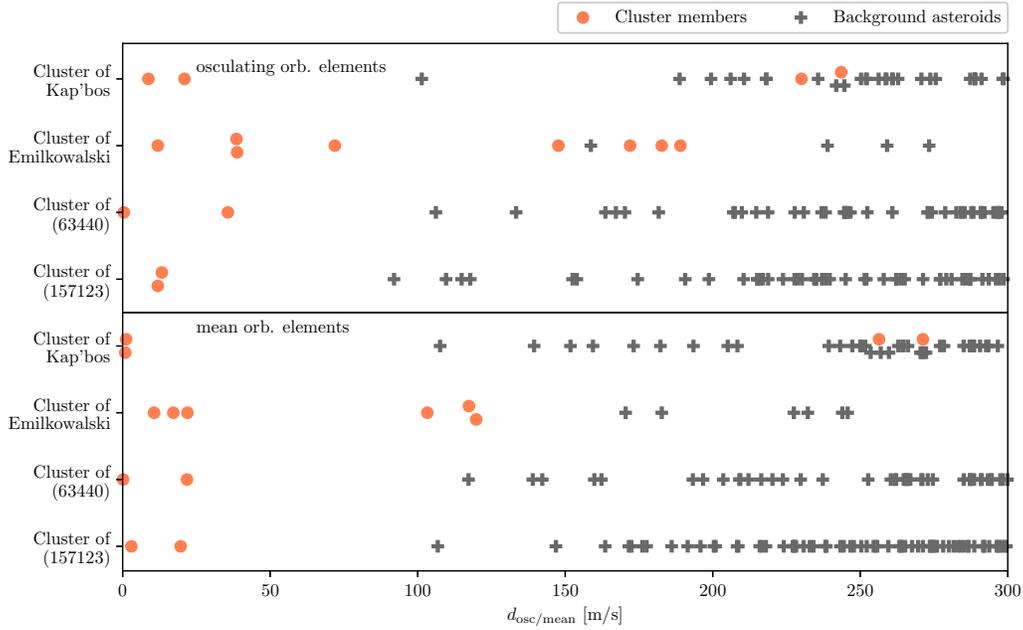}
\caption{The distribution of distances of asteroids around the primaries of the four studied clusters in both osculating (top) and mean (bottom) orbital elements. Mean orbital elements are only available for multi-opposition asteroids. Small offsets along the $y$ axis were applied to some points, where the cluster members' markers overlapped with other asteroid markers.}
\label{fig:closes_d}
\end{figure}

In each of the following four subsections we briefly summarize the identification history of given cluster, the identification procedure we used to found its new members and we discuss the distribution of the distances of the cluster members in the space of osculating and mean orbital elements. In Figure~\ref{fig:closes_d}, for each of the four clusters we show a distance distribution of all asteroids in vicinity to the primary up to  $d_{\mathrm{osc/mean}} = 300$ m/s, with the confirmed cluster members highlighted. It is notable that there is an apparent gap, especially in the mean orbital elements, between the cluster members and the background asteroids. This already provides a hint for statistical significance of the cluster that we later substantiate by calculation of the probability $p_1$ mentioned above. The exception is the cluster of Kab'bos, where the two distant members are mixed with the background asteroid population.

Further in each subsection, we present the results of our backward orbital integrations and the results of a test of nominal orbit integration with the Yarkovsky clones (described in Section~\ref{sec_4.1}).

\subsection{Cluster of (11842) Kap'bos}
\label{sec_4.1}
The two largest members of this cluster, (11842) Kap'bos and (228747) 2002~VH3 were identified as a significant pair by \citet{Pravec:2009} and its age was estimated to be $>$ 150 kyr \citep{Pravec:2010}. \citet{Pravec:2018} found that asteroid (436415) 2011~AW46 belongs to this pair as well, thus it is actually a cluster. We found two new members of this cluster by the extended search around the primary: (349108) 2007~GD18 and (445874) 2012~TS255. These two asteroids are much more distant from the primary that the two close secondaries above (see Table \ref{tab:11842}). In fact, they are the 10th and the 13th closest asteroids to the primary in the space of osculating orbital elements and the 19th and the 28th closest in the space of mean orbital elements\footnote{We used the \texttt{AstDyS-2} databases of osculating and mean orbital elements for numbered and multi-opposition asteroids, downloaded 2019-07-07.}, respectively. Therefore they were not discovered in \citet{Pravec:2018} where we limited the search to smaller distances. The probability of the secondary (228747) being only a close background asteroid (interloper) is $p_1 = 1.2 \times 10^{-11}$ with $R_0 = 168.7$ m/s (we could not apply the probability estimation for secondaries (445874) and (349108) since they are already mixing with the background asteroid population).

\begin{figure}[h!]
\centering\includegraphics[width=\linewidth]{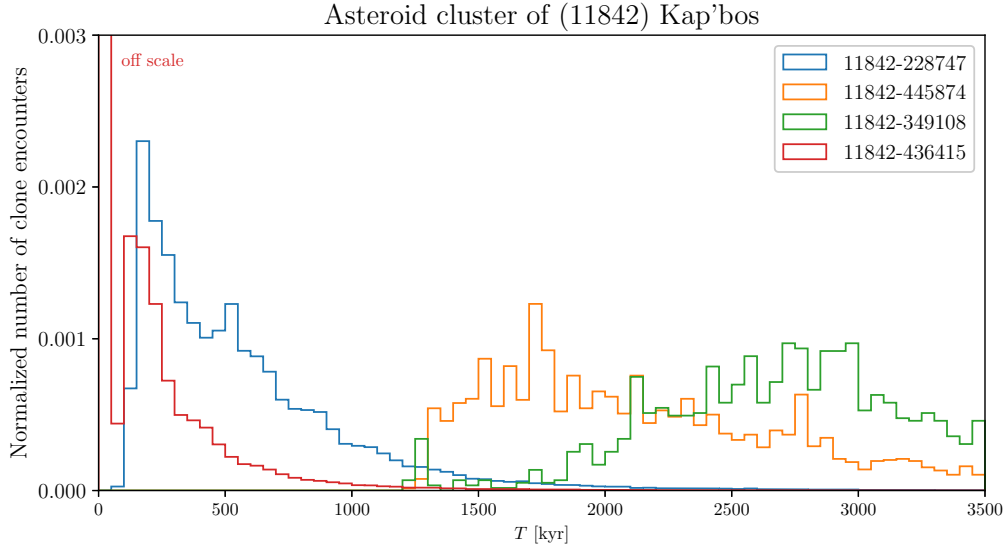}
\caption{Histogram of the orbital clone encounter times between the primary (11842) and the four secondaries of this cluster. We used limits $v_{rel} \le 2 v_{\mathrm{esc}}$ and $r_{\mathrm{rel}} \le 10 R_{\mathrm{Hill}}$ of the primary to filter slow and close clone encounters.}
\label{fig:11842a}
\end{figure}

For each of the four secondaries we performed backward orbital integrations and we plotted a histogram of times of close and slow clone encounters with the primary clones (Figure \ref{fig:11842a}). To overcome the difficulty of visualizing the distributions for highly different numbers of clone encounters -younger secondaries have typically many more recorded encounters than older secondaries- we normalized the histograms so that the sum of all bars of given distribution is equal to 1. In Table \ref{tab:11842} we give the estimated separation time for each secondary from the primary. The full description of the procedure is in Section~3. 

The estimated separation times of the secondaries from the primary appear to be divided into two groups with two members each. The secondaries (228747) and (436415) are very close to the primary and to each other with $d_{\mathrm{mean}} \sim 1$ m/s. Their separation times estimated by our backward orbital integrations are $465^{+917}_{-308}$ and $226^{+679}_{-127}$ kyr ago, respectively. We note that we also obtained many close and slow encounters between the orbital clones of (436415) and the primary at times about 15 kyr in the past. We believe that they do not indicate a real separation event, as all these encounters occurred at distances greater than $\sim4~ R_{\mathrm{Hill}}$, while the older encounters occurred at smaller relative distances. In similarly young asteroid pairs, the relative distances of clones during slow encounters are fractions of $R_{\mathrm{Hill}}$ \citep[e.g.,][]{Vokrouhlicky:2017b}. This suggest that it was just a close encounter between the two asteroids caused by a synodical cycle \citep[see][]{Zizka:2016} as previously suggested in \citet{Pravec:2018}. The other two secondaries, (349108) and (445874), are much more distant from the primary with $d_{\mathrm{mean}} \approx 256$ and 271 m/s, respectively. However, the relative distance $d_{\mathrm{mean}}$ between these two secondaries is only 18.6 m/s. The estimated separation times of these two secondaries from the primary are $>~1400$ kyr, which has only a very small overlap with the estimated ages of the two close secondaries.

We also performed a simple test of convergence of secular angles $\Omega \left( t \right)$ and $\varpi \left( t \right)$ of the nominal orbits of the cluster members with three Yarkovsky clones\footnote{Yarkovsky clones share the same initial orbital elements, but have different Yarkovsky effect strength acting on them.}. The three clones were assigned with zero, the maximum positive and the maximum negative Yarkovsky acceleration possible for the size and distance of the asteroid from the Sun. We then compared the time of $\Omega \left( t \right)$ and $\varpi \left( t \right)$ convergence (if it occurred) for all the nine combinations of the Yarkovsky clones between the primary and each secondary. The motivation for this exercise was to find the shortest possible time for the two orbits to become coplanar, which is a necessary requirement for slow encounters. A typical $\Delta \Omega$ and $\Delta \varpi$ for a young asteroid pair is $< 1^{\circ}$ at current epoch. The current $\Omega, \varpi$ differences between the secondaries (228747), (436415) and the primary are $< 0.3^{\circ}$, whereas the differences between the primary and the secondaries (349108), (445874) are $\Delta \Omega \sim 9^{\circ}$ and $11^{\circ}$, respectively, and  $\Delta \varpi \sim 14^{\circ}$ and $15^{\circ}$, respectively. For the secondaries (349108) and (445874) the shortest time for $\Omega \left( t \right)$ and $\varpi \left( t \right)$  convergences are 1000 and 1100 kyr, respectively, and therefore their younger close clone encounters at low relative velocities are highly improbable.

\begin{figure}[h!]
\centering\includegraphics[width=\linewidth]{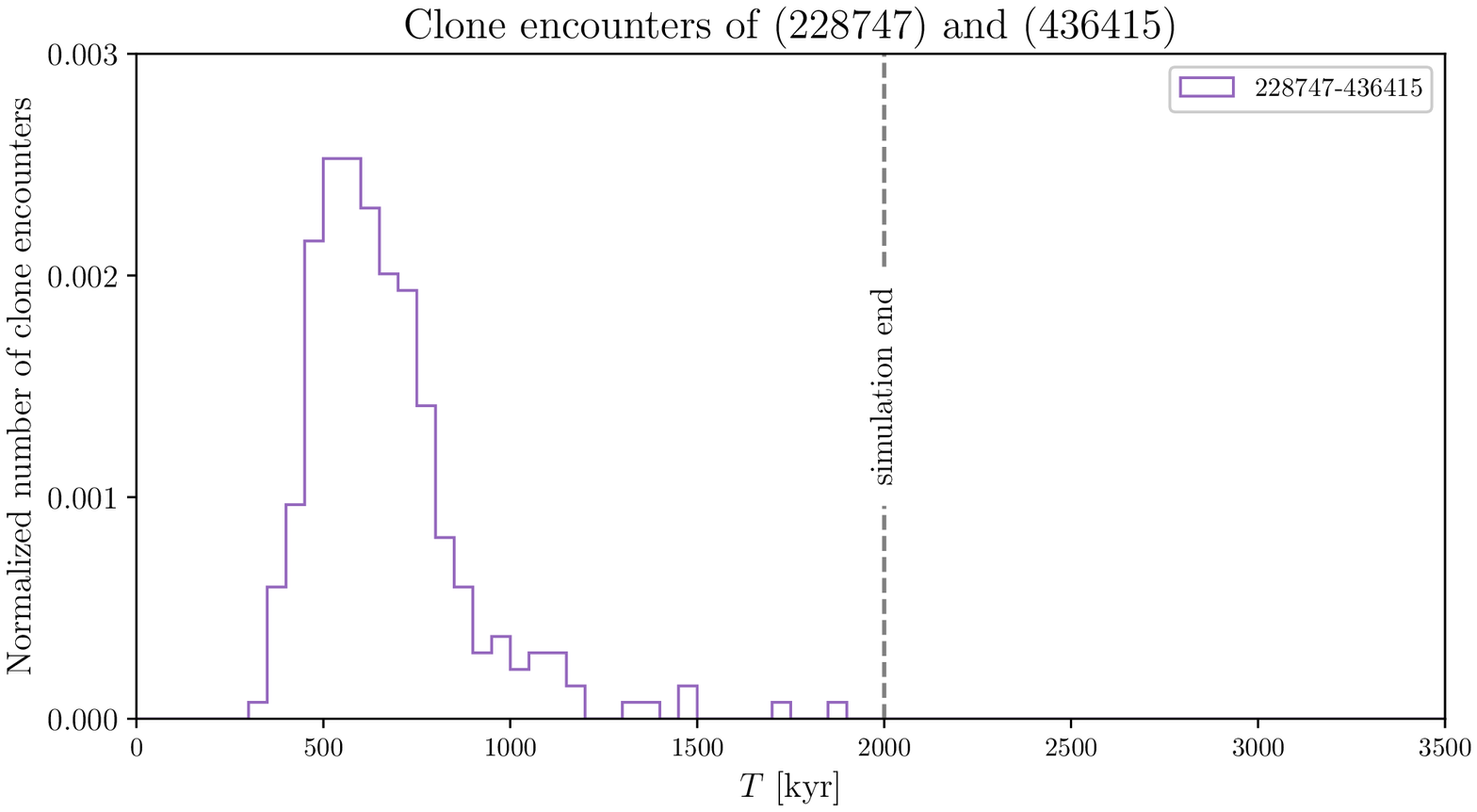}
\caption{Histogram of the orbital clone encounter times between the secondaries (228747) and (436415). We used limits $v_{rel} \le 2 v_{\mathrm{esc}}$ and $r_{\mathrm{rel}} \le 10 R_{\mathrm{Hill}}$ of (228747). The backward propagation of the clone orbits was stopped at 2 Myr (see the dashed line).}
\label{fig:11842_sec}
\end{figure}

Similarly to \citet{Carruba:2019} we searched for possible ``tertiary clusters'' or ``tertiary pairs'', which were formed by rotational fission of a secondary inside an existing cluster. In Figure \ref{fig:11842a} we see a significant overlap of the separation time distributions of the secondaries (228747) and (436415). The time distribution of clone encounters between these two secondaries is plotted in Figure \ref{fig:11842_sec}. We obtained a quite narrow time distribution of clone encounters with a hypothetical separation time of $626^{+462}_{-220}$ kyr ago. This result tells us that the disruption of a hypothetical parent secondary, leading to the formation of (228747) and (436415) cannot be ruled out as an alternative scenario, however their mutual distances $d_{\rm{osc}} = 29.6$ m/s and $d_{\rm{mean}} = 1.2$ m/s are higher than their distance to the primary Kap'bos. In any case, however, it is certain that these two secondaries are younger than the other two members of this cluster. We did not find any other clone encounters between other pairs of secondaries in this cluster.



\begin{table}
\begin{center}
\caption{Members of the asteroid cluster of (11842) Kap'bos with their absolute magnitudes $H$, distances $d_{\mathrm{osc/mean}}$ to the primary and estimated separation times $T_{\mathrm{sep}}$ in the past from the primary. In brackets is the ordinal number of given asteroid ordered by the  distance from the primary in given orbital elements.}
\renewcommand{\arraystretch}{1.2}
\begin{tabular}{ r l | c | c | c | c}
\multicolumn{2}{c |}{Asteroid} & H [mag]    & $d_{\mathrm{osc}}$ [m/s] & $d_{\mathrm{mean}}$ [m/s] & $T_{\mathrm{sep}}$ [kyr] \\
\hline
(11842)  & Kap'bos    & $14.42 \pm  0.03^a$   & -           & -           & - \\
(228747) & 2002~VH3   & $17.16 \pm  0.04^a$   & ~20.9 (2.)  & ~~1.1 (2.)  & $465^{+917}_{-308}$ \\
(445874) & 2012~TS255 & $17.9 $ 			  & 243.4 (13.) & 271.2 (28.) & $2017^{+1156}_{-623}$ \\
(349108) & 2007~GD18  & $18.0 $ 			  & 230.1 (10.) & 256.4 (19.) & $2708^{+656}_{-838}$ \\
(436415) & 2011~AW46  & $18.3 $ 		  	  & ~~8.7 (1.)  & ~~0.9 (1.)  & $226^{+679}_{-127}$ $^b$ \\
\end{tabular}

\label{tab:11842}
\end{center}
{\footnotesize $^a$From \citet{Pravec:2018}; the remaining $H$ values is from the \texttt{AstDyS-2} database.} \\
{\footnotesize $^b$See discussion of the age estimation in the third paragraph of Section 4.1.}

\end{table}

\subsection{Cluster of (14627) Emilkowalski}
A cluster of three asteroids (14627) Emilkowalski, (126761) 2002~DW10 and (224559) 2005~WU178 was found by \citet{Nesvorny:2006a}. The estimated age obtained by their backward orbital integrations was $\sim220$~kyr, however the authors noted that the perihelion convergence of (126761) with the primary (14627) was not perfect (see Fig. 2 of their paper). The fourth member of this cluster, (256124) 2006~UK337 was identified later and its backward orbital integrations showed consistency with the previously estimated age of this cluster. \citet{Pravec:2018} found three more multi-opposition members, (434002) 2000~SM320, 2014~UV143, (476673) 2008~TN44 and one probable single-opposition member 2009~VF107. The new multi-opposition secondaries were $\sim 5\times$ more distant in the space of mean orbital elements from the primary than the most distant previously known member (see Table \ref{tab:14627}). It is also notable that these new members are relatively close to each other with mutual $d_{\rm{mean}} \le 18$ m/s and they could be considered as a second core of this cluster \citep[see discussion in][]{Pravec:2018}. We found a new probable single-opposition member 2018~VB69, which is the second closest asteroid to the primary with $d_{\mathrm{osc}} \approx 31$ m/s, but due to the high uncertainties of its current orbital elements, we did not search for its close and slow encounters with the primary. For the most distant member in mean orbital elements (434002), we calculated $R_0 = 236.8$ m/s and $p_1 = 3.3 \times 10^{-2}$, which is still reasonably low. We also point out that the three most distant secondaries (434002), (476673) and 2014 UV143 are very close to each other, which substantially lowers the overall probability of each of these members to be an interloper.

\begin{figure}[h]
\centering\includegraphics[width=\linewidth]{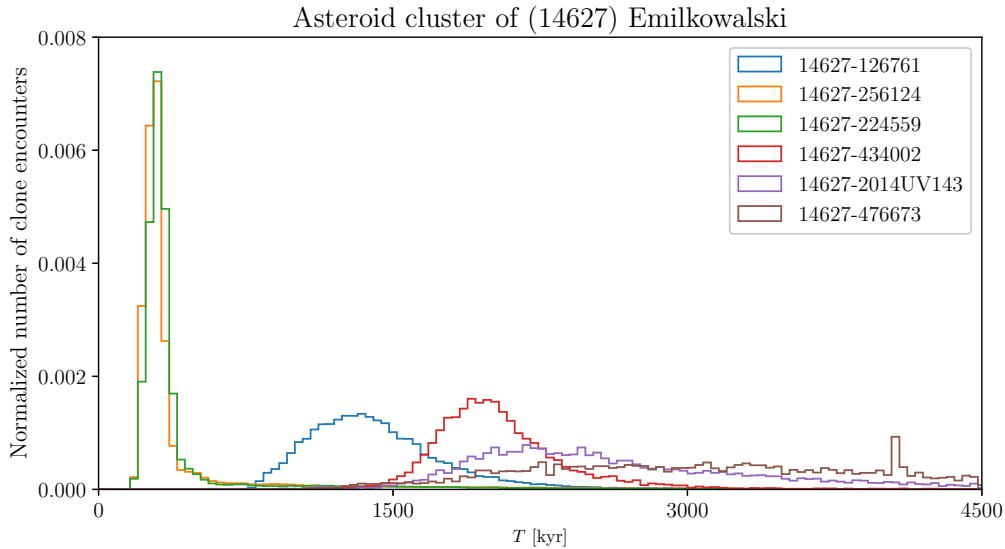}
\caption{Histogram of the orbital clone encounter times between the primary (14627) and the six multi-opposition secondaries of this cluster. We used limits $v_{\mathrm{rel}} \le 4 v_{\mathrm{esc}}$ and $r_{\mathrm{rel}} \le 15 R_{\mathrm{Hill}}$ of the primary.}
\label{fig:14627}
\end{figure}

From our backward orbital integrations we obtained a very nice distribution overlap of the clone encounters of the secondaries (224559) and (256124) with the primary (Figure \ref{fig:14627}). These two asteroids are the closest members to the primary and also the distance between them is quite small with the mutual $d_{\rm{mean}} = 6.6$ m/s. The estimated separation times for these two secondaries are $311^{+1183}_{-86}$ and $294^{+1452}_{-77}$ kyr ago, respectively. For the other four secondaries we obtained encounters with the primary at times $> 1$ Myr ago (Table \ref{tab:14627}) and there is little or no overlap between their distributions and the ones of the two young secondaries (224559) and (256124). The secondaries (434002), 2014 UV143 and (476673) have similarly placed leading edges of the age histograms at $\sim 1200$ kyr. The separation time distribution of the secondary (126761) is shifted towards younger ages by $\sim ~400$ kyr, but it still has no overlap with the dominant histogram peaks of the young secondaries (224559) and (256124). 

Our $\Omega \left( t \right)$ and $\varpi \left( t \right)$ convergence tests (see description in Section 4.1) for the cluster members suggest that asteroids (224559) and (256124) could not have slow encounters with the primary less than $\sim 200$ kyr ago, which is in agreement with the results of the backward integrations. The members of the second, older core of this cluster, (434002), 2014 UV143 and (476673) currently have $\Delta \Omega \sim 7^{\circ}$ and $\Delta \varpi \sim 1.5^{\circ}$ and they need at least 900, 700 and 1000 kyr, respectively, for their orbits to become coplanar with the orbit of the primary. Because of the very small difference of their secular orbital elements $\Delta \Omega \left( t=0 \right) \approx 0.3^{\circ}$ and  $\Delta \varpi \left( t=0 \right) \approx 1.3^{\circ}$ and the up to $3^{\circ}$ $\Delta \varpi$ amplitude oscillation we cannot set any time constraints for possible slow encounters of the secondary (126761) and the primary based on this two-element test.

\begin{figure}[h]
\centering\includegraphics[width=\linewidth]{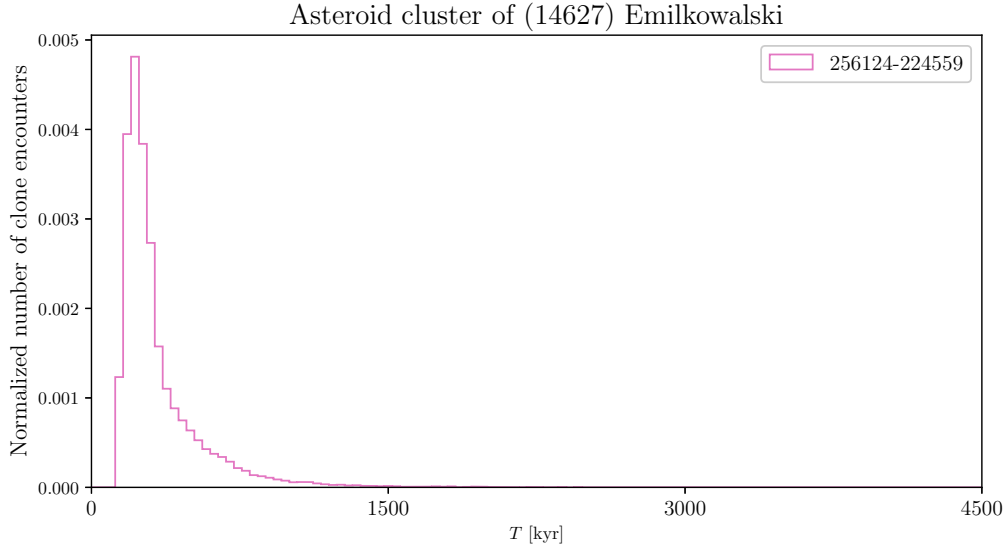}
\caption{Histogram of the orbital clone encounter times between the secondaries (256124) and (224559). We used limits $v_{\mathrm{rel}} \le 4 v_{\mathrm{esc}}$ and $r_{\mathrm{rel}} \le 15 R_{\mathrm{Hill}}$ of the larger secondary (256124).}
\label{fig:14627_sec}
\end{figure}

Similarly to the case of Kap'bos cluster, we see a clear overlap of encounter times distributions in Figure \ref{fig:14627} for the secondaries (256124) and (224559). We performed also a search for encounters between any two secondaries. In Figure \ref{fig:14627_sec} is plotted the time distribution of encounters between secondaries (256124) and (224559) with a hypothetical separation time of $264^{+509}_{-104}$ kyr ago. These two asteroids are even closer to each other in the space of mean orbital elements with $d_{\rm{mean}} = 6.6$ m/s, but we still cannot resolve whether they separated from each other or from the primary. We also found several encounters between the largest secondary (126761) and the two youngest secondaries (256124) and (224559) at times around 1300 and 1200 kyr, respectively. However, the number of clone encounters was much smaller (by a factor of $\sim1000$) than in the case of clone encounters of the Emilkowalski asteroid and any of these secondaries. Also the $d_{\rm{mean}}$ of (256124) and (224559) to the (126761) are more than $2\times$ and $3\times$ greater than to Emilkowalski, respectively. We did not find any other clone encounters between other pairs of secondaries.


\begin{table}
\begin{center}
\caption{Members of the asteroid cluster of (14627) Emilkowalski with their absolute magnitudes $H$, distances $d_{\mathrm{osc/mean}}$ to the primary and estimated separation times $T_{\mathrm{sep}}$ in the past from the primary. The single-opposition asteroids 2009~VF107 and 2018~VB69 do not have calculated mean orbital elements and we did not perform backward orbital integrations because of the large uncertainties of its orbital elements.}
\renewcommand{\arraystretch}{1.2}
\begin{tabular}{ r l | c | c | c | c}
\multicolumn{2}{c |}{Asteroid}   & H [mag]    & $d_{\mathrm{osc}}$ [m/s] & $d_{\mathrm{mean}}$ [m/s] & $T_{\mathrm{sep}}$ [kyr] \\
\hline
(14627)  & Emilkowalski & $13.61 \pm  0.06^a$ & -          & -          & - \\
(126761) & 2002~DW10    & $15.3			    $ & ~71.9 (4.) & ~22.0 (3.) & $1368^{+770}_{-414}$ \\
(256124) & 2006~UK337   & $15.9			    $ & ~11.9 (1.) & ~17.2 (2.) & $294^{+1452}_{-77}$ \\
(224559) & 2005~WU178   & $16.6			    $ & ~38.8 (3.) & ~10.7 (1.) & $311^{+1183}_{-86}$ \\
(434002) & 2000~SM320   & $16.9			    $ & 189.1 (9.) & 119.8 (6.) & $1991^{+724}_{-385}$ \\
         & 2014~UV143   & $17.5			    $ & 147.7 (5.) & 103.3 (4.) & $2470^{+1500}_{-750}$ \\
         & 2009~VF107   & $17.6			    $ & 172.0 (7.) &  -         &  -		    \\
(476673) & 2008~TN44    & $17.8			    $ & 182.7 (8.) & 117.4 (5.) & $3020^{+1232}_{-1340}$ \\
         & 2018~VB69    & $18.0^b		    $ & ~30.9 (2.) &  -         &  -		    \\

\end{tabular}
\label{tab:14627}
\end{center}
{\footnotesize $^a$From \citet{Pravec:2018}.} \\
{\footnotesize $^b$From \texttt{JPL Small-Body Database}. The remaining $H$ values is from the \texttt{AstDyS-2} database.}
\end{table}

\subsection{Cluster of (63440) 2001 MD30}
Asteroids (63440) 2001~MD30 and (331933) 2004~TV14 (located in the Hungaria region) were recognized as the tightest asteroid pair by \citet{Vokrouhlicky:2008b}. \citet{Pravec:2019} noted that asteroid 2008~VS46 also appeared to belong to this pair, but a detailed study of this possible cluster was beyond the scope of that study. We have verified its membership and thus these three asteroids are now classified as an asteroid cluster. The secondary (331933) is extremely close to the primary both in osculating and mean orbital elements (see Table \ref{tab:63440}). The secondary 2008~VS46 is somewhat more distant, but it is still the second closest asteroid to the primary in osculating as well as mean orbital elements.  We calculated that $R_0 = 172.7$ m/s and for the most distant secondary 2008~VS46 the probability $p_1 = 3.2\times10^{-5}$. 
 
\begin{figure}[h]
\centering\includegraphics[width=\linewidth]{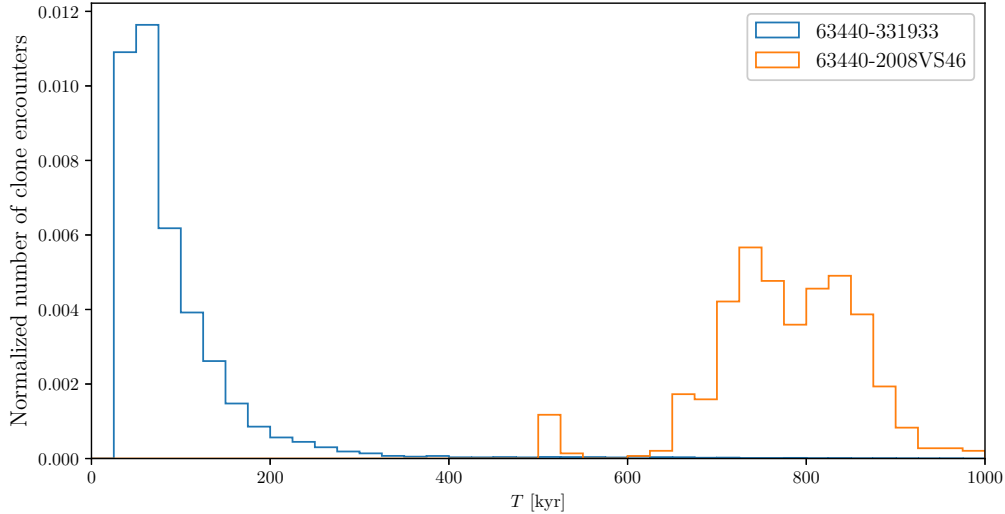}
\caption{Histogram of the orbital clone encounter times between the primary (63440) and its two secondaries. We plot the histogram of time encounters with $v_{\mathrm{rel}} \le 1 v_{\mathrm{esc}}$ and $r_{\mathrm{rel}} \le 10 R_{\mathrm{Hill}}$ for the secondary (331933) and $v_{\mathrm{rel}} \le 2 v_{\mathrm{esc}}$ and $r_{\mathrm{rel}} \le 15 R_{\mathrm{Hill}}$ for the secondary 2008~VS46 to partially compensate its higher initial orbital uncertainty, greater Yarkovsky effect strength uncertainty (due to its small size) and a higher clone dissipation at higher ages.}
\label{fig:63440}
\end{figure}

Our backward orbital integrations show two, clearly not overlapping time distributions of clone encounters for each of the two secondaries (see \ref{fig:63440}). As expected, the secondary (331933), which is very close  to the primary, has close and slow separation at lower ages, with the encounters time distribution peak around 70 kyr ago. The more distant secondary 2008~VS46 has an estimated age of $778^{+112}_{-119}$ kyr.

Because of the high similarity of the orbital elements of the secondary (331933) and the primary, especially in secular angles $\Delta \Omega \left( t=0 \right)~\approx~5~\times~10^{-4~\circ}$ and $\Delta \varpi \left( t=0 \right)~\approx~1.5~\times~10^{-2~\circ}$, we cannot put any constraint on the lowest possible age based on our secular angles convergence test. However, for the secondary 2008~VS46, we restrict that its separation from the primary occurred at least 270 kyr ago.

\begin{table}
\begin{center}
\caption{Members of the asteroid cluster of (63440) 2001~MD30 with their absolute magnitudes $H$, distances $d_{\mathrm{osc/mean}}$ to the primary and estimated separation times $T_{\mathrm{sep}}$ in the past from the primary.}
\renewcommand{\arraystretch}{1.2}
\begin{tabular}{ r l | c | c | c | c}
\multicolumn{2}{c |}{Asteroid}  & H [mag]    & $d_{\mathrm{osc}}$ [m/s] & $d_{\mathrm{mean}}$ [m/s] & $T_{\mathrm{sep}}$ [kyr] \\
\hline
(63440)  & 2001~MD30 & $15.63 \pm  0.13^a$ & -          & -          & - \\
(331933) & 2004~TV14 & $17.4			 $ & ~~0.4 (1.) & ~~0.1 (1.) & $68^{+151}_{-31}$ \\
         & 2008~VS46 & $19.2			 $ & ~35.7 (2.) & ~21.8 (2.) & $778^{+112}_{-119}$ \\
\end{tabular}
\label{tab:63440}
\end{center}
{\footnotesize $^a$From \citet{Pravec:2018}; the remaining $H$ values is from the \texttt{AstDyS-2} database.}
\end{table}

\subsection{Cluster of (157123) 2004 NW5}

This asteroid cluster was discovered by \citet{Pravec:2019} as a by-product of their search for asteroid pairs. It consists of the primary (157123) 2004~NW5 and two secondaries (385728) 2005~UG350 and 2002~QM97. The secondary 2002~QM97 is the closest asteroid to the primary with $d_{\mathrm{mean}} = 3.0$ m/s and the secondary (385728) is the second closest asteroid to the primary with $d_{\mathrm{mean}} = 19.7$ m/s. We did not find any new members of this cluster by our extended search around the primary. We calculated that $R_0 = 163.7$ m/s and for the most distant secondary 2008~VS46 the probability $p_1 = 2.5\times10^{-5}$. 

\begin{figure}[h!]
\centering\includegraphics[width=\linewidth]{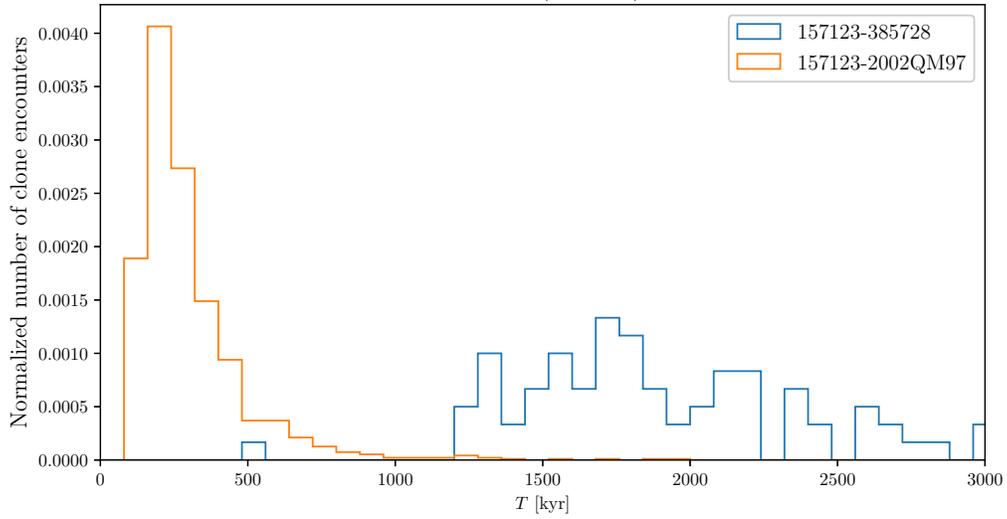}
\caption{Histogram of the orbital clone encounter times between the primary and the two secondaries of its cluster. We plot the histogram of encounters with $v_{\mathrm{rel}} \le 2 v_{\mathrm{esc}}$ and $r_{\mathrm{rel}} \le 10 R_{\mathrm{Hill}}$ limits for the secondary 2002~QM97 and $v_{\mathrm{rel}} \le 4 v_{\mathrm{esc}}$ and $r_{\mathrm{rel}} \le 15 R_{\mathrm{Hill}}$ for the secondary (385728) to partially compensate its more chaotic orbit and probable earlier separation time.}
\label{fig:157123}

\end{figure}

According to the \texttt{AstDyS-2} website, the Lyapunov characteristic exponent (quantifying dynamical chaos/predictability) for the primary is $\lambda = 26.87$~Myr$^{-1}$. This means that the distance between two initially close orbits will increase by factor of $e$ (Euler's number) in $\sim 37$~kyr. This chaoticity rapidly increases the uncertainty of our backward integrations further we go into the past. However, we were still able to obtain a reasonably well defined encounter time distribution for the secondary 2002~QM97 with an estimated separation time of $248^{+397}_{-114}$ kyr ago (Figure \ref{fig:157123} and Table \ref{tab:157123})\footnote{Our estimated age for 2002~QM97 is higher than the estimated age in \citet{Pravec:2019}, which was 146$^{+380}_{-88}$ kyr. This is because we used the somewhat tighter limits $r_{\mathrm{rel}} \leq 10 R_{\mathrm{Hill}}$ and $v_{\mathrm{rel}} \leq 2 v_{\mathrm{esc}}$ in this work, rather than the loosened limits $r_{\mathrm{rel}} \leq 15 R_{\mathrm{Hill}}$ and $v_{\mathrm{rel}} \leq 4 v_{\mathrm{esc}}$ used in \citet{Pravec:2019}.}. For the secondary (385728) we obtained a much worse time distribution of clone encounters despite having comparable uncertainties in the orbital elements, larger estimated size (meaning smaller range of possible Yarkovsky effect strength) and smaller Lyapunov characteristic exponent than 2002~QM97. We obtained times of clone encounters ranging from 1200 kyr up to 3000 kyr ago, where our simulation ended. (We also found a few clone encounters around 520 kyr ago, but we consider them insignificant.) We believe that the broad shape of the clone encounter times distribution of the secondary (385728) is a result of its higher age and the resulting orbit chaoticity that goes with it.

An analysis of the evolution of $\Delta \Omega$ and $\Delta \varpi$ of the Yarkovsky clones is less sensitive to the orbital chaoticity than the close and slow clone encounter search. We estimated that the minimum time required for the orbits of the primary and (385728) to become coplanar is around 420 kyr. Even though this lower constraint is located at the trailing edge of the histogram of clone encounters for 2002~QM97 (meaning that a single separation event for this cluster is not entirely ruled out), an explanation including two separate events should be considered. Considering the unusually small difference of the primary's and the largest secondary's absolute magnitudes ($\Delta H = 0.7$), it is also possible that (157123) may not be the real primary of this cluster, but only the largest secondary, while the real primary still has to be found. Nevertheless, we have not found any other suitable candidate for the real primary yet, which could be because of the relatively high orbital chaoticity in the given region of the main belt. 

\begin{table}
\begin{center}
\caption{Members of the asteroid cluster of (157123) 2004~NW5 with their absolute magnitudes $H$, distances $d_{\mathrm{osc/mean}}$ to the primary and estimated separation times $T_{\mathrm{sep}}$ in the past from the primary.}
\renewcommand{\arraystretch}{1.2}
\begin{tabular}{ r l | c | c | c | c}
\multicolumn{2}{c |}{Asteroid}   & H [mag]    & $d_{\mathrm{osc}}$ [m/s] & $d_{\mathrm{mean}}$ [m/s] & $T_{\mathrm{sep}}$ [kyr] \\
\hline
(157123) & 2004~NW5   & $16.93 \pm  0.07^a$ & -          & -          & - \\
(385728) & 2005~UG350 & $17.6			  $ & ~13.3 (2.) & ~19.7 (2.) & $1792^{+922}_{-496}$ \\
         & 2002~QM97  & $18.6			  $ & ~11.9 (1.) & ~~3.0 (1.) & $248^{+397}_{-114}$ \\
\end{tabular}
\label{tab:157123}
\end{center}
{\footnotesize $^a$From \citet{Pravec:2018}; the remaining $H$ values is from the \texttt{AstDyS-2} database.}
\end{table}

\clearpage

\section{Possible cascade disruption mechanism}
\citet{Pravec:2018} proposed, from their analysis of cluster primary rotations and mass ratios, that most asteroid clusters were formed by rotational fission of parent rubble pile asteroids. They discussed two possible mechanisms for formation of $\geq 2$ secondaries. One is called ``secondary fission'' and it was proposed by \citet{Jacobson:2011}. It is a rotational fission of the secondary induced via spin-orbit coupling that occurs during the temporary chaotic binary stage of the initial phase of an asteroid pair evolution. From the newly formed chaotic ternary system, one or both secondaries may escape if the system has a positive free energy. The escape happens very rapidly (typically in $< 1$ yr after the fission event). Before escaping the secondaries may undergo further secondary fission event(s) creating a more complex system with three or more secondaries. The second discussed formation mechanism was proposed by \citet{Vokrouhlicky:2017a}. They suggested that a swarm of small fragments can be the result of a cratering event from an impact of a small projectile onto a nearly critically rotating primary. \citet{Pravec:2018} suggested that this mechanism may be more probable for a cluster with many members, such as the Datura family with estimated $\sim300$ members with sizes $> 200$ m. However, neither of these two proposed cluster formation mechanisms explains the cascade fission seen in the four studied clusters in this work. 


We consider that a possible mechanism for a cascade formation of a cluster are two (or more) rotational fission events of a parent body at different times. The proposed scenario is following. The parent (rubble pile) asteroid undergoes a rotational fission (after being spun-up to its critical frequency by the YORP effect), the newly formed secondary/ies escape/s and the rotation frequency of the remnant parent asteroid (we call it ``intermediate parent'' in following) is decreased (rotation slowed down). The intermediate parent is then spun up by the YORP effect and it reaches the critical rotation frequency again and another fission event occurs. 

We test the hypothesis with following model. We assumed that the shapes of the current primaries and its parent bodies are a prolate spheroid with $a_\mathrm{p}, b_\mathrm{p}, c_\mathrm{p}$ being its principal semi-axes, with $a_\mathrm{p} \geq b_\mathrm{p} = c_\mathrm{p}$ and the $c_\mathrm{p}$ axis being the rotational axis of the body. The equatorial elongation $a_\mathrm{p}/b_\mathrm{p}$ was estimated from the observed lightcurve amplitude $A$ of the primary using the relation\footnote{This relation is valid for zero solar phase angle and for an equator-on viewing aspect. In a general case these two conditions are not met. However, the non-zero solar phase angle increases an observed amplitude, whereas the tilted rotational axis decreases it. This means that these two effects work in opposite directions and we consider this equation to be a reasonable approximation for the asteroid equatorial elongation for our purpose.} $A = 2.5 \log \left(a_\mathrm{p}/b_\mathrm{p}\right)$. We started with a parent body with the mass equal to the sum of masses of all known cluster members with an assumed bulk density $\rho$ and geometric albedo $p_{\mathrm{v}}$. The parent asteroid was rotating at its critical frequency\footnote{The critical frequency $f_{\mathrm{crit}}$ (number of rotations per unit of time) is  when the centrifugal force  is equal to the gravitational force at some point of the asteroid's surface. In the case of a prolate spheroid with the rotational axis identical with the semi-minor axis $c_{\mathrm{p}}$, this happens at the longest ends of the prolate body and the critical frequency is $f_{\mathrm{crit}} = \sqrt{ G \rho \left( e_\mathrm{p}^2 -1 \right) \left[ 2 e_\mathrm{p} + \ln \left( \frac{1-e_\mathrm{p}}{1+e_\mathrm{p}} \right) \right] / \left( 2 \pi e_\mathrm{p}^3 \right)}$, where $e_\mathrm{p} \equiv \sqrt{1 - b_\mathrm{p}^2 / a_\mathrm{p}^2}$ \citep{Richardson:2005}.} $f_{\mathrm{crit}}$, then it split up and one or more secondaries (first generation) escaped from the longest ends of the body at an escape velocity of the remaining primary body (intermediate parent). We assumed that the energy needed for all the secondaries to escape was transferred from the rotation of the parent body and we calculated the new rotational frequency of the intermediate parent. Like in \citet{Pravec:2008}, we scaled the strength of the YORP effect from \citet{Capek:2004} to the size, bulk density and distance of the parent asteroid from the Sun to suit our modeled asteroids and estimated the rotational acceleration $\dot{f}$ caused by the YORP effect. Then we calculated the time needed for the intermediate parent to reach $f_{\mathrm{crit}}$ again, so it could undergo another fission event. We then calculated the rotational frequency of the resulting primary after the second rotational fission and compared it with its current observed primary rotation period. The results of these calculations are discussed for the individual clusters in  following subsections. We note that this model is very simplified and its results should be treated as the first rough approximation, intended as a test whether the multi-fission disruption scenario is possible. Because of the lack of detailed information for the parameters of the cluster members and their parents, we are forced to use several approximations and assumptions, specifically: we assumed that we know all members of a given cluster; the rotational energy of the secondaries can be  neglected; the observed amplitude represents the asteroid's elongation; rotational axes of the primary, the intermediate parent and the grand-parent asteroids are perpendicular to the orbital plane (meaning the most effective YORP acceleration); the shapes of the primary, the intermediate parent and the grand-parent asteroids are the same, only their sizes are different.

\subsection{Cluster of (11842) Kap'bos}
Kap'bos has $a = 2.25$ au, $e = 0.095$ and $i = 3.69^{\circ}$, it is located in the Flora family and is probably an S type \citep{Popescu:2018, Pravec:2018}, therefore we assumed its geometric albedo $p_{\mathrm{V}} = 0.2$ and the bulk density $\rho = 2$ g/cm$^3$. Its absolute magnitude is $H = 14.42 \pm 0.03$ with the mean lightcurve amplitude $A = 0.13$ mag. Its rotational period is $P = 3.68578 \pm 0.00009$ h (all from Pravec et al., 2018).

The estimated critical rotation frequency for a parent body of this cluster is $f_{\mathrm{crit}} = 9.79 $ rot/d and the rotational frequency decreased after the first fission event and the escape of the oldest secondaries (445874) and (349108) to $f_1 =  9.50$ rot/d. With the estimated $\dot{f} = 0.134 $ rot/d/Myr of its intermediate parent, the time needed for its spin-up to the critical frequency $f_{\mathrm{crit}}$ for the second fission event is about 2.16 Myr, which is in good agreement with the estimated separation times of about 2 Myr obtained from our backward integrations. The evaluated rotational frequency after the second fission and escape of the young secondaries (228747) and (436415) is $f_2 = 9.23$ rot/d, which is faster by about 42\% than the current observed period of the primary. This difference could be due to a few possible factors; ($i$) Incomplete membership of the cluster - undiscovered young secondary/ies, whose escape from the primary would require additional energy transfer from its rotation, which would slow down the primary. ($ii$) The intermediate parent could be more elongated than the current primary; more elongated bodies have lower $f_{\rm{crit}}$. ($iii$) The real bulk density of the asteroid could be lower than assumed. Note that $f_{\rm{crit}}$ depends on the bulk density as $f_{\rm{crit}} \sim \sqrt{\rho}$, therefore lower density means lower critical rotational frequency.

\subsection{Cluster of (14627) Emilkowalski}
(14627) Emilkowalski is located in the central Main Belt ($a = 2.60$ au, $e = 0.15$ and $i = 17.75^{\circ}$). \citet{Veres:2015} classified this asteroid to be a D type. Since there have not been obtained reliable density measurements for D type asteroids yet \citep{Carry:2012}, we adopted the density value $\rho = 1 $ g/cm$^{3}$ as a compromise between primitive C type asteroids (with typical densities 1.5 g/cm$^{3}$) and comets (with typical densities about 0.5 g/cm$^{3}$). We used its refined albedo $p_{\mathrm{V}} = 0.13$ derived by \citet{Pravec:2018}\footnote{The derived albedo by \citet{Pravec:2018} is inconsistent with the typical values for D type asteroids \citep{Burbine:2016}. We plan to verify the taxonomy classification of the primary by obtaining its spectra.}. The absolute magnitude of the primary is $H = 13.61 \pm 0.06$ with the mean lightcurve amplitude $A = 0.67$ mag and its rotational period is $P = 11.1313 \pm 0.0009$ h \citep{Pravec:2018}.

There exists a young (still forming) dust band in the main asteroid belt observed by \textit{IRAS}\footnote{\textit{InfraRed Astronomical Satellite}} that several authors \citep[e.g.,][]{Vokrouhlicky:2008a, Espy:2015} associated with the Emilkowalski cluster. The estimated age of this dust band is significantly less than 1~Myr \citep{Espy:2015} and it is similar to the time of the latest disruption event of the Emilkowalski cluster, which is about 300 kyr before present. \citet{Espy:2015} estimated the amount of particles in the dust band to be equivalent to a $\sim3-4$ meters deep layer of regolith on the surface of a $\sim 8$ km diameter parent body. They also estimated that the ejection velocity of the dust particles was a few times the escape velocity from the parent body of the Emilkowalski cluster to provide a good fit to the inclination dispersion of the observed band. We took the dust band into account in the following test by estimating its total mass and calculating the energy needed for its escape from the primary at various relative velocities.

The estimated critical frequency of a parent asteroid of the Emilkowalski cluster is $f_{\mathrm{crit}} = 5.46$ rot/d and after the first fission event and the escape of the secondaries (126761), (434002), 2014 UV143, (476673) and 2009~VF107\footnote{We excluded the probable, one-opposition member 2018~VB69 from this test, because it is unclear when it separated from the primary.} the rotational frequency decreases to $f_1 = 3.09$ rot/d. With the estimated $\dot{f} = 0.061$ rot/d/Myr it would take the intermediate parent  about 38.7~Myr to reach the $f_{\mathrm{crit}}$ again to make the second fission possible. This estimated time is significantly longer than the $\sim2$ Myr estimated from the backward orbital integrations (see discussion below). Once the $f_{\mathrm{crit}}$ was reached again and the secondaries (256124), (224559) together with all the dust particles were ejected with the escape velocity, the rotational frequency decreased to $f_2 = 4.302$ rot/d, which is about 2 times faster than the current observed rotational frequency of the primary. However, if we allow the dust particles to be ejected at velocity $\sim3.8\times$ the escape velocity from the primary, then the resulting rotational frequency matches the currently observed one. We remind that this higher ejection velocity of the dust particles is also required by the \citet{Espy:2015} study of the dust band.

While the unusual slow rotation of the primary ($\sim2\times$ slower than the second slowest primary rotation of all known clusters) can be explained by including the ejected dust, the main issue here is the time needed for the second fission to happen. Even with the assumption of the most effective YORP effect configuration, the estimated time is $\sim20\times$ longer that the age estimated by the backward orbital integrations. Possible factors affecting it are : ($i$) The grand-parent asteroid was less elongated than the intermediate parent, which would mean higher $f_{\mathrm{crit}}$ before the first fission and the intermediate parent would keep more rotational energy after the separation event. ($ii$) The intermediate parent was more elongated than the current primary is. This would mean that $f_{\mathrm{crit}}$ of the intermediate parent would be lower and could be reached faster. ($iii$) The real bulk density is lower than assumed. For instance, with $\rho = 0.5$ g/cm$^3$, the time needed for spin-up of the intermediate parent to $f_{\mathrm{crit}}$ is reduced by 65\% to$~13.7$ Myr. ($iv$) Another mechanism (such as non-gravitational spin-up by jets on an active asteroid/cometary nucleus) or a collision is involved in the formation process of this unusual asteroid cluster. The observed properties (the D type classification, the dust band presence) together with the results of our simulation suggest that the Emilkowalski may in fact be a cometary nucleus\footnote{It might be originally a trans-Neptunian object that was transported to the main belt during the planet reconfiguration 4 billion years ago \citep{Vokrouhlicky:2016b}.}. We plan to do a more thorough study of this cluster in the future. Specifically, we plan to confirm its taxonomic type, obtain more photometric data needed for determining its shape as well as spin pole and perform more detailed backward orbital integrations.

\subsection{Cluster of (63440) 2001 MD30}
This cluster lies in the Hungaria asteroid group ($a = 1.94$ au, $e = 0.09$ and $i = 19.99^{\circ}$). \citet{Polishook:2014} and \citet{Pravec:2019} classified its primary to be an X/E type asteroid. We adopted following physical parameters for our test: $\rho = 2$ g/cm$^3$ and $p_{\mathrm{V}} = 0.4$ \citep{Warner:2009}. Its absolute magnitude is $H = 15.63 \pm 0.13$, the lightcurve amplitude $A = 0.15$ mag and the rotational period is $P = 3.2969 \pm 0.0002$ h \citep{Pravec:2019}. 

For the parent asteroid of this cluster, we calculated the $f_{\mathrm{crit}} = 9.71$ rot/d and after the first fission event and an escape of the older secondary 2008 VS46, the rotation slows down to $f_1 = 9.58$ rot/d. This is a rather small change due to the small size of 2008 VS46. This means that with the estimated $\dot{f} = 1.062$ rot/d/Myr, it takes only about 120 kyr for the intermediate parent to reach $f_{\mathrm{crit}}$ and to undergo the second fission. After the second fission and the escape of (331933), the rotation slows down to $f_2 =   7.86 $ rot/d, which is faster by 8\% than the currently observed primary rotation rate. This suggests that there is no other (yet undiscovered) young secondary of a size comparable to (331933) in this cluster. The formally short estimated time needed for the intermediate parent to reach $f_{\mathrm{crit}}$ again is not an issue and it can be explained by ($i$) more (yet undiscovered) secondaries separated together with the older secondary 2008 VS46, or ($ii$) the intermediate parent was not in the optimal configuration, which led to a lower $\dot{f}$, thus longer time was needed to reach the $f_{\mathrm{crit}}$ again.


\subsection{Cluster of (157123) 2004 NW5}
The primary (157123) is located in the inner Main Belt ($a = 2.31$ au, $e = 0.24$ and $i = 4.13^{\circ}$). It is probably a S type \citep{Pravec:2019}, therefore we assumed $p_V = 0.2$ and $\rho = 2$ g/cm$^3$. Its absolute magnitude is $H = 16.93 \pm 0.07$, the mean lightcurve amplitude $A = 0.65$ mag and the rotation period is $P = 3.5858 \pm 0.0005$ h \citep{Pravec:2019}.

The estimated critical rotational frequency is $f_{\mathrm{crit}} = 7.80$ rot/d. We note that formally there was not enough energy in the rotation of the parent asteroid for the large secondary (385728) to escape after the first fission; in fact its escape formally requires at least $~166\%$ of the rotational energy of the parent body at $f_{\mathrm{crit}}$. The secondary (385728) holds about 1/4 of the total mass of this cluster. With the calculated difference of the equivalent secondary magnitude and the primary magnitude of $\Delta H \sim 0.5$ and the observed primary rotation period, this cluster lies outside the allowed limits predicted by the theory of formation by rotational fission \citep[see Fig. 14 in][]{Pravec:2018}. Nevertheless, the fission of the intermediate parent asteroid is possible in our model. After the latest fission and escape of 2002 QM97, the rotational frequency was lowered to the frequency $f_2 = 4.98$ rot/d, which is slower by $\sim26\%$ than the current observed primary rotation frequency. This is in a good agreement if we consider the typical catalog uncertainty of the $H$ estimations for objects for which we don't have more precise measurements (both secondaries in this case) and the fact that the real bulk density of the cluster members can be a little different from the assumed value.

The issue of the escape of the secondary (385728) may have several causes, such as: $(i)$ Another mechanisms is involved in the formation of this cluster, like for the clusters of Hobson and Mandragora that are also located outside the allowed limits by the rotational fission theory \citep{Pravec:2018}. ($ii$) The rough value of $H$ taken from the orbit catalog for the secondary (385728) is in error and the real $H$ value is higher, meaning that the asteroid is smaller and less energy is required for its escape. ($iii$) The asteroid (157123) is not in fact the real primary of this cluster, but only the largest secondary (see discussion in Section 4.4.). However, the probability of the asteroid cluster of (157123) being only a chance coincidence of three genetically unrelated asteroids located close to each other in the space of orbital elements is extremely low, so it is a securely identified cluster.


\section{Summary and conclusions}

We found two new members of the cluster of Kap'bos, (349108) and (445874), which separated from the primary significantly earlier in the past than the two younger secondaries (228747) and (436415). The estimated time difference between these two secondary separation events is about $\sim 2$ Myr, which is very close to the time needed (2.16 Myr) for the intermediate-parent of this cluster to reach its critical rotational frequency by the YORP effect, per our model. If we consider the uncertainties of the secondary age estimates, the assumed physical parameters and the simplicity of our model, the proposed theory of two fission events of the primary invoked  by the YORP effect is consistent with our data for the Kap'bos cluster.

For the cluster of Emilkowalski, we found one new single-opposition asteroid 2018~VB69 that is most likely a member of this cluster. We will confirm its membership (together with the other single-opposition asteroid 2009~VF107 that was found before) when its orbit's accuracy is improved by new astrometric observations in the future. The result of our backward orbital integrations clearly indicate at least two separation events in the last 5 Myr. The shortest possible time needed for the second fission estimated by our model is $\sim 38.7$ Myr formally, which is almost $20\times$ longer than the estimated time between the two disruption events suggested by the backward orbital integrations. So, the rotational acceleration by the YORP effect only cannot explain the formation of this cluster. The apparent association of the cluster with the observed dust band at $i \approx 17^{\circ}$, the proposed primary's D type taxonomic classification and the relatively slow rotation of the primary make this cluster a very interesting case, which deserves more attention in the future. In particular, data on the primary's shape, its bulk density and rotational axis orientation (of the primary as well the secondaries) would significantly advance our understanding of this cluster.

For the cluster of (63440), we confirmed that the asteroid 2008~VS46 is related to the previously known pair (63440) - (331933). Our backward orbital integrations showed a clear gap between the time distributions of the slow and close clone encounters for the two secondaries. The estimated time difference between these two secondary separation time distributions is several times larger than the estimated time of $\sim120$ kyr needed for the YORP-induced spin-up to $f_{\mathrm{crit}}$ after the first fission event obtained from our model. This means that the formation mechanism of this cluster can be explained by two rotational fission events of the same asteroid and the YORP acceleration is sufficient. Our model allows for an existence of another possible cluster member of similar size as the secondary 2008~VS46, which may be discovered in the future.

For the cluster of (157123), we reproduced the backward orbital integrations from \citet{Pravec:2019} and confirmed the membership of the secondaries (385728) and 2002~QM97 and their separations at different times. Despite the broad time distribution for clone encounters of (385728) with the primary, a clear gap is visible between it and the time distribution of 2002~QM97. The predicted rotation frequency of the primary is reasonably close to the current observed value, which supports the idea that 2002~QM97 separated from the (157123). However, our model formally does not allow an escape of the large secondary (385728), raising the question about accuracy of the estimated cluster parameters, or whether the secondary (385728) really separated from (157123). In other words, we are not certain that the asteroid (157123) is the real primary of this cluster.

Finally, we looked at whether there is not possibly some common property for these four clusters. Two of them have probably S type primaries, one X/E type and one probable D type primary. One of the cluster is located in the Hungaria asteroid group ($a = 1.94$ au), one in the Flora family ($a = 2.25$ au, being a region of stable orbits), one in the inner part of the main asteroid belt ($a = 2.31$ au, located close to a resonance) and one in the central part of the main belt ($a = 2.60$ au). Two of the clusters have 3 known members, one has 5 known members and one has 7 to 9 known members (2 being single-opposition asteroids). There does not seem to be an obvious common property of the four studied clusters. It looks like cascade disruptions may occur in about any asteroid cluster.

\section*{Acknowledgments}
We are grateful to the reviewers, Valerio Carruba and Bojan Novakovi\'{c}, for their useful comments and suggestions that led us to improve several points in this paper. Petr Fatka was supported by the Charles University, project GA UK No. 842218. This work was supported by the Grant Agency of the Czech Republic, Grant 17-00774S. Computational resources were provided by the CESNET LM2015042 and the CERIT Scientific Cloud LM2015085, provided under the programme ``Projects of Large Research, Development, and Innovations Infrastructures''.

\section*{References}

\bibliographystyle{apalike} 
\bibliography{references}

\newpage

\end{document}